\documentstyle[12pt]{article}
\newcommand{\m}{\medbreak}
\newcommand{\no}{\noindent}
\newcommand{\EQ}{\begin{equation}}
\newcommand{\eq}{\end{equation}}
\def\EQA{\begin{eqnarray}}
\def\eqa{\end{eqnarray}}
\newcommand{\ALL}{\mbox{$A_{LL}\ $}}
\newcommand{\ALLPV}{\mbox{$A_{LL}^{PV}\ $}}
\newcommand{\ALLPVBAR}{\mbox{${\bar A}_{LL}^{PV}\ $}}
\newcommand{\AL}{\mbox{$A_{L}\ $}}

\begin{document}
\begin{titlepage}
\vspace{0.2in}
\vspace*{1.5cm}
\begin{center}
{\large \bf Testing various polarized parton distributions at RHIC
\\} 
\vspace*{0.8cm}
{\bf Jacques Soffer}{$^1$} and {\bf Jean Marc Virey}{$^{2,a}$}  \\ \vspace*{1cm}
{$^1$}Centre de Physique Th\'eorique$^{\ast}$, C.N.R.S. - Luminy,
Case 907\\
F-13288 Marseille Cedex 9, France\\ \vspace*{0.2cm}
{$^2$}Centre de Physique Th\'eorique and
Universit\'e de Provence, Marseille, France\\
\vspace*{1.8cm}
{\bf Abstract \\}
\end{center}
A very promising spin physics programme will be soon on the way at the BNL
Relativistic Heavy Ion Collider (RHIC). By studying the spin asymmetries for various
processes (single photon, single-jet and $W^{\pm}$ production), we will compare the
different predictions obtained using some sets of polarized parton distributions,
available in the recent literature. We will put some emphasize on the analysis of the
anticipated errors, given the event rates expected from this high luminosity new
machine and the current acceptance for the detector systems at RHIC.

\vfill
\begin{flushleft}
PACS Numbers : 

\m\no
May 1997\\
CPT-97/P.3495\\
\m\no
anonymous ftp or gopher : cpt.univ-mrs.fr

------------------------------------\\
$^{\ast}$Unit\'e Propre de Recherche 7061

{$^a$} Moniteur CIES and allocataire MESR \\
E-mail : Soffer@cpt.univ-mrs.fr ; Virey@cpt.univ-mrs.fr
\end{flushleft}
\end{titlepage}

\section{Introduction}
\indent
\m
Serious progress have been made over the last ten years or so, in our understanding
of the spin structure of the nucleon. This is partly due to the discovery in 1987, by
the EMC experiment \cite{Ashman}, of the violation of the Ellis-Jaffe sum rule 
\cite{EllisJaffe}. This result was obtained by measuring the proton polarized
structure function $g_1^p (x,Q^2)$, in polarized deep-inelastic scattering with a
polarized amonia target ($NH_3$). It led to an unexpected situation, that is, only a
small fraction of the proton spin is carried by quarks, which is at variance with
the naive quark-parton model. The EMC result has motivated several experiments, in
three different laboratories CERN , DESY and SLAC \cite{Amsterdam}, to undertake some
systematic determinations of $g_1^{p,n} (x,Q^2)$, for proton and neutron by means of
measurements using a polarized lepton beam and several different polarized targets
(hydrogen, deuterium, helium-3). 
The main goal of these experimental programmes was
to confirm the EMC result on $g_1^p (x,Q^2)$ and to measure the neutron structure
function $g_1^n (x,Q^2)$, in order to test the fundamental Bjorken sum rule
\cite{Bjorken}. It turns out that these fixed target experiments allow to cover only
a rather limited kinematic domain that is, 0.005  $< x <$ 0.7, with the
corresponding average $Q^2$, $<Q^2>$ between 2 $\, GeV^2$ and 10 $\, GeV^2$. Needless
to recall that the low $x$ region, which is not easily accessible and therefore loosely
determined, is very crucial for testing accurately these sum rules. In addition it is
not possible to make a flavor separation to isolate the contribution of each quark to
the nucleon spin.

On the theoretical side, the surprising EMC result has also stimulated an intense
activity,  in particular its possible interpretation in terms of the so-called axial
anomaly which involves the gluon polarization $\Delta G(x,Q^2)$
\cite{Anselmino,Cheng,Lampe}. This is an interesting physical quantity in itself,
but it can only be determined indirectly in polarized deep-inelastic scattering. More
precisely the contribution of the gluon polarization to the Ellis-Jaffe sum rules is
ambiguous and, as we will see below, much cleaner information will be extracted at
the RHIC facility from collisions with polarized proton beams. Let us also mention
that several important results in higher order perturbative QCD corrections have
been obtained recently \cite{Anselmino,Cheng,Lampe}, which will allow, in the
future, very stringent tests of the QCD spin sector, so far essentially unexplored.

In order to describe the most recent experimental results on $g_1^{p,n,d} (x,Q^2)$,
several phenomenological models have been proposed and correspondingly, one can find
in the recent literature various sets of parametrization for the polarized
parton distributions. In spite of the constant progress in the accuracy of the data,
the extraction of these distributions, especially for sea quarks and gluons, remains
fairly ambiguous and controversial. We will select some of them and recall their main
features in Section 2, where we also give some basic kinematics and the definitions of
the spin observables we will study below. In Section 3 we will consider direct photon
production. In terms of perturbative QCD, it is a simple process whose double helicity
asymmetry is particularly sensitive to $\Delta G (x,Q^2)$, as we will illustrate by
means of a useful approximate expression. Section 4 is devoted to single-jet
production which has a higher complexity than the previous process but whose generous
cross section will allow a very accurate determination of the double helicity
asymmetry. We will consider in Section 5 the production of $W^{\pm}$ gauge bosons and
parity violating asymmetries with either one or two polarized beams. Their
measurements will provide a good calibration of the quark and antiquark polarizations
and also the $u$ and $d$ flavor separation. We will comment on the possible benefit
one can get by measuring them simultaneously in proton-proton and proton-neutron
collisions. We will give our concluding remarks in Section 6.

\section{Spin observables and various parametrizations}
\m
\indent

Fundamental interactions at short distances which are probed in high energy hadronic
collisions, involve hard scattering of quarks, antiquarks and gluons. Let us consider
the general hadronic reaction
\EQ
a\, +\, b \rightarrow \, c + \, X
\eq
\noindent
where $c$, in the cases we will consider below, is either a photon, a $W^{\pm}$
or a single-jet. In the hard scattering kinematic region, the cross section describing
(1) reads in the QCD parton model, provided factorization holds, as
\EQ
d\sigma (a+b \rightarrow c+X) \, =\, 
\sum_{ij}{1\over1+\delta_
{ij}}\int dx_adx_b\biggl[f^{(a)}_i\bigl(x_a,Q^2\bigl)
f^{(b)}_j\bigl(x_b,Q^2\bigl)\,d\hat{\sigma}^{ij}\; +\; (i\leftrightarrow j)
\biggl].
\eq
The summation runs over all contributing parton configurations, the
$f(x,Q^2)$'s are the parton distributions, directly extracted from deep-inelastic
scattering for quarks and antiquarks and indirectly for gluons.
d${\hat \sigma}_{ij}$ is the cross section for the interaction of two partons
$i$ and $j$ which can be calculated perturbatively, some of which , at the lowest
order, are given in ref.\cite{BRST}.
If we consider the reaction (1) with {\it {both}} initial hadrons, $a$ and $b$
longitudinally polarized, one useful observable is the {\it{double}}
helicity asymmetry \ALL defined as
\EQ
A_{LL} ={d\sigma_{a(+)b(+)}-d\sigma_{a(+)b(-)}\over 
d\sigma_{a(+)b(+)}+d\sigma_{a(+)b(-)}}\; ,
\eq \noindent
when we assume parity conservation, i.e.
d$\sigma_{a(\lambda)b(\lambda')}$ = d$\sigma_{a(-\lambda)b(-\lambda')}$.
Its explicit expression, assuming factorization, is given by
\EQ
A_{LL}d\sigma\, =\, 
\sum_{ij}{1\over1+\delta_{ij}}\int
dx_adx_b\biggl[\Delta f^{(a)}_i\bigl(x_a,Q^2\bigl) \Delta
f^{(b)}_j\bigl(x_b,Q^2\bigl) \hat {a}_{LL}^{ij}
d\hat{\sigma}^{ij}+(i\leftrightarrow j)\biggl]\; ,
\eq \noindent
where d$\sigma$ is given by eq.(2) and $\hat {a}_{LL}^{ij}$ denotes the corresponding subprocess
double asymmetry for initial partons $i$ and $j$. The $\Delta f$'s 
are defined as
\EQ
\Delta f(x,Q^2)\; =\; f_+(x,Q^2) \, +\, f_-(x,Q^2)
\eq \noindent
where $f_{\pm}$ are the parton distributions in a polarized hadron with helicity
either parallel (+) or antiparallel (-) to the parent hadron helicity.
Recall that the unpolarized distributions are $f\, = \, f_+ \, + \,f_-$ and
$\Delta f$ measures how much the parton $f$ "remembers" the parent hadron
helicity.
If the subprocess involves parity violating interactions, one can consider another
interesting observable which requires only {\it {one}} initial hadron polarized,
that is the {\it {single}} helicity asymmetry \AL, defined as
\EQ
A_{L} = {d\sigma_{a(-)}-d\sigma_{a(+)}\over 
d\sigma_{a(-)}+d\sigma_{a(+)}}\, . 
\eq 
In addition, if both $a$ and $b$ are polarized one can also have two
{\it {double}} helicity parity violating asymmetries defined as
\EQ
A_{LL}^{PV} ={d\sigma_{a(-)b(-)}-d\sigma_{a(+)b(+)}\over 
d\sigma_{a(-)b(-)}+d\sigma_{a(+)b(+)}}
\;\;\;\;\; and \;\;\;\;\;
\bar A_{LL}^{PV} ={d\sigma_{a(-)b(+)}-d\sigma_{a(+)b(-)}\over 
d\sigma_{a(-)b(+)}+d\sigma_{a(+)b(-)}} \; ,
\eq \noindent
which can be simply related to \AL as we will see in section 5, where we will give
their explicit expressions for $W^{\pm}$ production. This completes the presentation
of the spin observables and let us now turn to a short discussion on the selected
sets of polarized parton distributions.
\m
We first consider the set of polarized parton densities which has been proposed in
ref.\cite{BS95}. The quark and antiquark densities are parametrized in terms of 
Fermi-Dirac distributions with only {\it {nine}} free parameters. Simple
assumptions allow to relate unpolarized and polarized quark distributions and the free
parameters are determined from the very accurate data on deep inelastic neutrino
scattering at $Q^2 = 3\,$ GeV$^2$. For the gluon distributions one uses a
Bose-Einstein expression which is given with no additional free parameters. This
allows a straightforward DGLAP $Q^2$ evolution which leads to excellent predictions
for $F_2(x,Q^2)$ up to the $x,Q^2$ kinematic range accessible at HERA. These quark
densities give also a fair description of the low $Q^2$ structure functions
$g_1^{p,n}(x,Q^2)$. However, in this approach $\Delta G(x,Q^2)$ at 
$Q^2 = 3\,$ GeV$^2$ is not known and two possibilities have been proposed, namely the
{\it {soft}} and the {\it {hard}} gluon polarization which are shown
in Fig.1a.

There is another standard method for obtaining the polarized parton distributions,
which is more generally used in the literature. It is based on the direct analysis 
of $g_1^{p,n}(x,Q^2)$, independently of the unpolarized structure functions. 
A very simple \linebreak
parametrization for the quark densities, of the type
$x^{\alpha}(1-x)^{\beta}$, has been proposed in ref.\cite{BouBuc}. It was found that
for a correct description of the data, in particular the very recent accurate neutron
data \cite{HughesWarsaw} from SLAC E154 at $Q^2 = 5\,$ GeV$^2$,
one needs a rather substantial $\Delta G(x,Q^2)$, having also a simple
expression as above, which is displayed in Fig.1a.

In ref.\cite{GS96} the parametrization is based on more complicated expressions
of the type $x^{\alpha}(1-x)^{\beta}(1+\gamma {\sqrt{x}} + \rho x)$ in a similar way
as what was used by Martin-Roberts-Stirling to describe the unpolarized quark
distributions \cite{MRS}. Due to our poor knowledge of $\Delta G(x,Q^2)$, they
parametrize it in the same way and they explore various possibilities with three different
choices of $\gamma_G$ and $\rho_G$. They conclude that three scenarios A, B and C,
shown in Fig.1b are equally acceptable and therefore $\Delta G(x,Q^2)$ is largely
undetermined. However their analysis, which is prior to the SLAC E154 neutron data,
does not include it. This is also the case for the parametrization proposed in 
ref.\cite{GRSV}, where the polarized distributions are constructed from the
unpolarized ones, the so called GRV densities \cite{GRV} by means of a simple
multiplicative factor $x^a(1-x)^b$. Needless to recall that the free parameters are
such as to respect the fundamental positivity constraints down to the low resolution
scale of this approach namely $Q^2 = 0.23$ GeV$^2$. In this case, one gets a gluon
polarization, shown in Fig.1b, which is half way between the soft and the hard 
$\Delta G$ of ref.\cite{BS95}.

In addition to these four sets of polarized parton distributions we just briefly
discussed, many other different choices have bee presented in the literature
\cite{ChengLiu,AltaBall}. It appears that from the present available data, the
valence quark distributions are fairly well determined and rather consistent with
each other, unlike for the sea quarks (or antiquarks) and for gluons as we have seen
above.

Measurements of polarized deep inelastic lepton nucleon scattering yield direct
information on the spin asymmetry
\EQ
A_{1}^{N}\, \sim \, {g_1^N(x,Q^2) \over F_1^N(x,Q^2)}
\;\;\;\;
(N=p,n,d)\; ,
\eq \noindent
where $F_1^N(x,Q^2)$ is simply related to the unpolarized structure function
$F_2^N(x,Q^2)$. All data for $Q^2>1\,$ GeV$^2$ are consistent with a flat $Q^2$
behaviour of $A_1^N(x,Q^2)$ which reflects the fact that the scaling violations
of $g_1^N(x,Q^2)$ and $F_1^N(x,Q^2)$ are not so different
and not distinguishable within experimental errors. The next-to-leading
or leading order QCD evolutions leads to $A_1^N(x,Q^2)$ almost independent of $Q^2$
in accordance with experimental observation. To illustrate this fact we show
in Fig.2 the results for the proton case $A_1^p(x,Q^2)$ from ref.\cite{GRSV}.
This interesting property will be used in the next section on direct photon
production.
\section{Direct photon production}
\m
\indent

The cross section for direct photon production on $pp$ collisions at high
$p_T$ is considered as one of the cleanest probe of the unpolarized gluon
distribution $G(x,Q^2)$. This is partly due to the fact that the photon originates in
the hard scattering subprocess and is detected without undergoing fragmentation.
Moreover in $pp$ collisions the quark-gluon Compton subprocess
$qG \rightarrow q\gamma$ dominates largely and the quark-antiquark annihilation subprocess
$q{\bar q} \rightarrow G \gamma$ can be neglected. Consequently the double helicity asymmetry
$A_{LL}^{\gamma}$ (see eq.(4)), which involves in this case only one subprocess,
becomes particularly simple to calculate and is expected to be strongly sensitive to
the sign and magnitude of $\Delta G(x,Q^2)$. For the Compton subprocess, 
$\hat a_{LL}$ whose expression at the lowest order is given in ref.\cite{BRST},
is always positive and such that 
$\hat a_{LL}({\hat \theta}_{cm} = 90^{\circ})\, =\, 3/5$ where ${\hat \theta}_{cm}$
is the center of mass (c.m.) angle in the subprocess.

Before we proceed, we would like to present an approximate expression of
$A_{LL}^{\gamma}$ valid at pseudo-rapidity $\eta = 0$, where the cross section reaches
its largest values. To a good approximation we can write
\EQ
A_{LL}^{\gamma}(\eta=0) \; =\; {3 \over 5} \; {\Delta G (x_T,Q^2)
\over G (x_T,Q^2)} \, A_1^p (x_T,Q^2)\; ,
\eq \noindent
where $x_T = 2 p_T/\sqrt s$, $p_T$ being the transverse momentum of the outgoing
photon and $\sqrt s$ the c.m. energy of the reaction $pp \rightarrow \gamma X$.
In eq.(9) one should take $Q^2 = p_T^2$ and $A_1^p$ is the deep-inelastic spin
asymmetry we discussed above (see eq.(8)). This expression shows clearly the
dependence of $A_{LL}^{\gamma}$ on $\Delta G$ and, in order to test its validity, we
have compared the exact calculation of $A_{LL}^{\gamma}$, using eqs.(2) and
(4) also with $Q^2 = p_T^2$, to the above approximation. The results are depicted
in Fig.3a for two sets of polarized parton distributions and show the usefulness of
this simplified expression (9). $A_{LL}^{\gamma}$ increases with $p_T$ due to the
rapid growth of $A_1^p$ with $x$ and follows the sign of $\Delta G$ and its magnitude
which is larger in the case of BS-b than for GRSV (see Figs. 1a,b).

In Fig. 3b we show a more complete comparison of the results we obtained for
the $p_T^{\gamma}$ distribution of $A_{LL}^{\gamma}(\eta=0)$, using the different
sets of polarized parton distributions we have discussed in the previous section, with
a leading order $Q^2$ evolution. Actually we find that the smallest predictions
correspond to the sets ref.\cite{GS96} and ref.\cite{GRSV} which have the smallest
$\Delta G(x)/G(x)$. The predictions differ substantially at large $p_T$, which
corresponds to the region, say around $x=0.4$ or so, where the distributions
$\Delta G(x)$ have rather different shapes. We have also indicated the expected
statistical errors based on an integrated luminosity $L = 800$ pb$^{-1}$ at
$\sqrt s = 500\,$ GeV, for three months running time. We have evaluated the event
rates in the pseudo-rapidity gap $-1.5 < \eta < 1.5$, assuming a detector efficiency of
100\% and for a $p_T^{\gamma}$ acceptance $\Delta p_T^{\gamma} = 5\,$ GeV/c.
We see that up to $p_T^{\gamma} = 50\,$ GeV/c or so, $A_{LL}^{\gamma}$ will be 
determined with an error less than 5\% which therefore will allow to distinguish
between these different possible $\Delta G(x)$. For very large $p_T^{\gamma}$,
the event rate drops too much to provide any sensitivity in the determination
of $\Delta G(x)$. In Fig. 4 we display the pseudo-rapidity distribution 
of $A_{LL}^{\gamma}(\eta)$ at a fixed $p_T^{\gamma}$ value.
It shows a systematic increase of $A_{LL}^{\gamma}$ for higher
$\eta$, already observed in ref.\cite{BouGuiSof}, which is partly due to the cross
section fall off away from $\eta = 0$. This smaller event rate leads also to
larger statistical errors, but this measurement is worth performing because it is
obviously a non-trivial QCD test.
\m
Let us now give a short discussion of these results and how they can be compared
with other recent calculations. First, one may be concerned about higher order QCD
corrections, to our dominant subprocess ${\vec q}\,{\vec g} \rightarrow q \,\gamma$.
The corresponding K-factors have been calculated \cite{Contogouris,GorVol}\footnote{
The results obtained in ref.\cite{GorVol}, which are slighty different from those of 
ref.\cite{Contogouris}, are based on the choice of a consistent scheme.}
and they always
exceed 1, so they rescale up separately all spin-dependent cross sections.
However this effect tends to be marginal on $A_{LL}^{\gamma}$ which is a ratio
of cross sections. Next we should consider the effect of a $Q^2$ evolution of
the polarized parton distributions which were done here at the leading order
(LO) while in ref.\cite{JaffeSaito} they took the next-to-leading order (NLO) version.
It is known that there is almost no difference for the valence ($u$ and $d$) 
quark distributions, but this is not the case for the gluons. Actually at the NLO, 
the maximum of $\Delta G(x)/G(x)$ is shifted towards smaller $x$ values 
and therefore the rise of $A_{LL}^{\gamma}$ is sharper. This is precisely what we see
by comparing the GS sets in Fig.3b with Fig.4 of ref.\cite{JaffeSaito}, which also
uses ref.\cite{GS96}. Finally there is another aspect to be mentioned.
At collider energies the outgoing photon is very often produced inside an hadronic 
jet, whose energy has to be restricted by isolation cuts in order to improve
the detection accuracy. This is a standard procedure experimentally but not at all
straightforward theoretically. It has been studied at the NLO with a  Monte-Carlo
method \cite{Gordon} and it was found that the effects of isolation cuts
are rather important for unpolarized cross sections, but not so much for 
$A_{LL}^{\gamma}$ and they decrease with increasing energy. Actually they become
minute at $\sqrt s = 500\,$ GeV and the predictions one finds in 
ref.\cite{Gordon} are almost identical to those of
ref.\cite{JaffeSaito} where isolation cuts were ignored.

\section{Single jet production}
\m
\indent

Inclusive jet production is also a physics area where one can learn a lot about parton
densities and, considering the vast amount of unpolarized existing data, it has been
regarded as an important QCD testing ground. Event rates are much larger than  for
prompt photon production, but there is a drawback because many subprocesses are
involved, unlike in the previous case. In principle one should take into account
gluon-gluon ($GG$), gluon-quark ($Gq$) and quark-quark ($qq$) scatterings. Although
these subprocesses cross sections are not so much different, after convolution
with the appropriate parton densities (see eq.(2)), they lead to very distinct
contributions to the hadronic spin-average cross section.

In the spin dependent case where no data exist so far, the same ingredients can be
used to test earlier determinations of the polarized parton densities and to check
the consistency of our picture of the nucleon spin. A first attempt in connection
with RHIC was proposed in ref.\cite{BouGuiSof}, where various aspects of jet physics
were discussed. Here we will restrict ourselves to the double helicity asymmetry
$A_{LL}^{jet}$ for single jet production and in order to clarify the interpretation
of our results below, let us recall some simple dynamical features. In the very
low $p_T^{jet}$ region, say $p_T^{jet} \sim 10$ GeV/c or so, $GG$ scattering dominates
by far, but its contribution drops down very rapidly with increasing $p_T^{jet}$.
In the medium $p_T^{jet}$ range, say 20 GeV/c $< p_T^{jet} < 80$ GeV/c or so, 
$Gq$ scattering dominates and then decreases for large $p_T^{jet}$, to be overcome by
$qq$ scattering. Of course these are rough qualitative considerations and accurate
numerical estimates for the relative fractions of these different
contributions depend strongly on the parton densities one uses.

Let us first look at $A_{LL}^{jet}(\eta=0)$ and from the above discussion we see 
that one does not expect an approximate expression similar to eq.(9). However in the
medium $p_T^{jet}$ range where $Gq$ scattering dominates, $A_{LL}^{jet}(\eta=0)$
should have a trend similar to $A_{LL}^{\gamma}(\eta=0)$, with perhaps a magnitude
reduced by a factor two, since about half of the jet cross section is due to $GG$ and
$qq$ scatterings. This is what we see approximately in Fig. 5, where we present the
numerical results for $A_{LL}^{jet}(\eta=0)$ at ${\sqrt s} = 500$ GeV, which should
be compared to Fig.3b. We have also indicated the statistical errors which are
extremely small in this case, because of the huge event rates. The $\eta$
distribution of $A_{LL}^{jet}(\eta)$ for $ p_T^{jet} = 60$ GeV/c is shown
in Fig. 6 and it is either flat or decreasing for increasing $\eta$.
This is at variance with the prompt photon case (see Fig.4), because although the
unpolarized jet cross section decreases away from $\eta = 0$
\footnote{Actually this theoretical prediction has never been checked
experimentally since the single-jet cross sections are always given 
for $\eta=0$ assuming a given pseudo-rapidity cut, 
usually $|\eta|<0.7$ or more.}, 
the numerator of $A_{LL}^{jet}(\eta)$ is sensitive to the behaviour of the different
contributions $GG$, $Gq$ and $qq$. We find that while $Gq$ increases with
$\eta$, both $GG$ and $qq$ are decreasing and therefore the precise behaviour of
$A_{LL}^{jet}(\eta)$ depends strongly on their relative sizes.
The statistical errors, assuming a realistic
$\Delta \eta = 0.2$ per bin, will be small enough to allow the
extraction of the correct trend from future data. Finally we recall that the
predictions of ref.\cite{JaffeSaito} are somehow different from ours, which is
certainly due to the use of different sets of polarized parton distributions.
As already noted in ref.\cite{JaffeSaito}, the sets of ref.\cite{BS95} and
ref.\cite{GS96} have different flavor decomposition and, as we will see in the next
section, $W^{\pm}$ production is certainly a very promising way to improve our
knowledge on the $u,d$ flavor separation.

\section{$W^{\pm}$ production}
\m
\indent

Let us first consider, for the reaction $pp \rightarrow W^{\pm}X$, the parity violating single
helicity asymmetry $A_L$ defined in eq.(6). In the Standard Model, the $W$ gauge
boson is a purely left-handed object and this asymmetry reads simply for
$W^{\pm}$ production 
\EQ
A_L^{W^+}(y)\, =\, \frac{\Delta u(x_a,M^2_W)\, {\bar d}(x_b,M^2_W)\, -\,
(u \leftrightarrow {\bar d})}{u(x_a,M^2_W)\, {\bar d}(x_b,M^2_W)\, +\,
(u \leftrightarrow {\bar d})}\; ,
\eq \noindent
assuming the proton $a$ is polarized. Here we have $x_a = {\sqrt \tau}e^y$, 
$x_b = {\sqrt \tau}e^{-y}$ and $\tau = M_W^2/s$. For $W^{-}$ production the quark
flavors are interchanged ($u \leftrightarrow d$). The calculation of these
asymmetries is therefore very simple and the results are presented in 
Figs.7a,b at ${\sqrt s} = 500$ GeV, for different sets of distributions.
As first noticed in ref.\cite{BS93}, the general trend of $A_L$ can be easily
understood as follows : at $y=0$ one has
\EQ
A_L^{W^+}\, =\, \frac{1}{2}\left( \frac{\Delta u}{u}\, -\, \frac{\Delta \bar d}
{\bar d} \right)\;\;\;\;\;\;\; and \;\;\;\;\;\;\; A_L^{W^-}\, =\,
\frac{1}{2}\left( \frac{\Delta d}{d}\, -\, \frac{\Delta \bar u}
{\bar u} \right)\; ,
\eq \noindent
evaluated at $x=M_W/{\sqrt s}=0.164$, for $y=-1$ one has
\EQ
A_L^{W^+}\, \sim\,  -\, \frac{\Delta \bar d}
{\bar d} \;\;\;\;\;\;\; and \;\;\;\;\;\;\; A_L^{W^-}\, \sim\,
 -\, \frac{\Delta \bar u}
{\bar u} \; ,
\eq \noindent
evaluated at $x=0.059$ and for $y=+1$ one has
\EQ
A_L^{W^+}\, \sim\, \frac{\Delta u}{u}\, \;\;\;\;\;\;\; and 
\;\;\;\;\;\;\; A_L^{W^-}\, \sim \,
\frac{\Delta d}{d}\; ,
\eq \noindent
evaluated at $x=0.435$. Therefore these measurements will allow a fairly clean 
flavor separation, both for quarks and antiquarks, for some interesting ranges of
$x-$values. We see in Fig.7a that $A_L^{W^+}$, which is driven by the $u$ and 
$\bar d$ polarizations, leads to similar predictions for all cases. This is mainly
due to our knowledge of $\Delta u/u$, except for $x\geq 0.3$ where it comes out to
be larger for BS-h \cite{BS95}. In Fig.7b, for $A_L^{W^-}$ which is sensitive to the
$d$ and $\bar u$ polarizations, we see that the various predictions leads to the same
general trend, with some difference in magnitude due to a large uncertainty in the
determination of $\Delta d/d$. Also BS-h \cite{BS95} has assumed a larger negative
$\Delta {\bar u} /u$ which is reflected in the behaviour near $y=-1$. The statistical
errors have been calculated with a rapidity resolution $\Delta y = 0.2$ and taking into
account only the events from the leptonic decay modes. They are smaller for $W^+$
production which has larger event rates.
\m
Let us now turn to the parity violating double helicity asymmetries
\ALLPV and \ALLPVBAR defined in eq.(7). Their explicit expressions are given
in ref.\cite{BS93} and we recall that in $pp$ collisions they have the 
following symmetry properties
\EQ
A_{LL}^{PV}(y) = A_{LL}^{PV}(-y) 
\;\;\;\;\; and \;\;\;\;\;
\bar A_{LL}^{PV}(y) = - \bar A_{LL}^{PV}(-y).
\eq
A priori these asymmetries are independent observables, but if one makes the
reasonable assumption $\Delta u \Delta \bar d \ll u \bar d$ for all $x$, one gets the
following relations
\EQ
A_{LL}^{PV}(y)\, = \,A_{L}(y)+A_{L}(-y) 
\;\;\;\;\;\; , \;\;\;\;\;\;
\bar A_{LL}^{PV}(y) \,=\, A_{L}(y)-A_{L}(-y).
\eq
Therefore these double helicity asymmetries do not contain any
additional information, but they help to emphasize the differences among the four
predictions for $A_L^{W^+}$ and $A_L^{W^-}$, as seen in Fig.8a,b for \ALLPV. From an
experimental viewpoint, they might be also more useful, if one does not have such a
small rapidity resolutions.
\m
Finally let us consider the realistic possibility of having proton-neutron collisions
at RHIC, which can be achieved with either deuteron or helium-3 beams. The parity
violating single helicity asymmetry $A_L$ for $W^+$ production in $pn$ collisions
reads \cite{BS94}
\EQ
A_L({\vec p}n \rightarrow W^+\, ;y)\, =\, \frac{\Delta u(x_a,M^2_W)\, {\bar u}(x_b,M^2_W)\, -\,
\Delta {\bar d}(x_a,M^2_W)\, d(x_b,M^2_W)}{u(x_a,M^2_W)\, {\bar u}(x_b,M^2_W)\, +\,
{\bar d}(x_a,M^2_W)\, d(x_b,M^2_W)}\; ,
\eq \noindent
assuming the proton is polarized. Of course one has a similar expression for $W^-$
production. The double helicity asymmetries are also related to the sum or the
difference of two $A_L$ but this time for \ALLPV, instead of eq.(15), we 
have the sum
\EQ
A_{LL}^{PV}({\vec p}{\vec n} \rightarrow W^+\, ;y)\, = \,
A_{L}({\vec p}n \rightarrow W^+\, ;y)\, +\, A_{L}({\vec p}n \rightarrow W^-\, ;-y)
\eq \no
and therefore the following symmetry property
\EQ
A_{LL}^{PV}({\vec p}{\vec n} \rightarrow W^+\, ;y)\, = \,
A_{LL}^{PV}({\vec p}{\vec n} \rightarrow W^-\, ;-y)
\eq

Similarly, we find for \ALLPVBAR that eq.(17) holds, with the difference instead of
the sum, and the symmetry property
\EQ
{\bar A}_{LL}^{PV}({\vec p}{\vec n} \rightarrow W^+\, ;y)\, = \,-\,
{\bar A}_{LL}^{PV}({\vec p}{\vec n} \rightarrow W^-\, ;-y)
\eq

Therefore eq.(17) allows us to calculate \ALLPV, even if one cannot have polarized
neutrons. We show in Figs.9a,b the four predictions as in the previous cases
\footnote{We realize that since neutron beams are not directly available, but they
require deuteron or helium-3 beams, ${\sqrt s} = 500$ GeV is
obviously too high to be accessible at
RHIC.}.

\section{Concluding remarks} 
\m
\indent

  In this paper we have briefly reviewed the specific properties of some
sets of polarized parton distributions which have been considered in the
recent literature, to describe polarized deep-inelastic scattering data.
We have stress the importance of the future polarized $pp$ collider at the
BNL-RHIC, for studing these distributions, in a kinematic range so far
inexplored and to provide the first serious tests of the QCD spin sector.

We have seen that because of the high luminosity of the machine, we will
be able to pin down the magnitude and the sign of $\Delta G(x,Q^2)$ which
is, at present, the less well-known distribution. This will allow to
verify directly one possible interpretation of the observed violation
of the Ellis-Jaffe sum rules \cite{EllisJaffe} for proton and neutron,
in terms of the axial anomaly. For this particular topic, one requires
detailed measurements of the double helicity asymmetry $A_{LL}$ for
single jet and direct photon production. These will be fairly accurately
determined as we have dicussed from the small size of the expected
statistical errors. 

The copious production of $W^\pm$ gauge bosons
at $\sqrt{s} = 500$ GeV will also provide an extremely valuable source
of information on these distributions and in particular will help us
to perform a clean flavor separation between $u$ and $d$ quarks and
similarly for antiquarks $\bar u$ and $\bar d$. This can be achieved
from the measurement of the single helicity asymmetry $A_L$, but we have
also discussed some interesting features of the double helicity parity
violating asymmetries \ALLPV and \ALLPVBAR which are simply related to $A_L$.
Although $W^\pm$ production will be easier in $pp$ collisions, we have
also considered some interesting aspects of the parity violating
asymmetries in the plausible situation where we might have $pn$ collisions.


\vspace*{3cm}
\no {\bf Acknowledgments}

\m \no
We thank C. Bourrely and P. Taxil for very helpfull discussions,
G. Bunce and M. J. Tannenbaum for valuable informations,
and all the participants of the RSC Annual Meeting held in Marseille.
 This work has been partially supported
by the EC contract CHRX-CT94-0579.
 

%

\newpage
{\bf Figure captions}
\bigbreak
\no
{\bf Fig. 1a} The gluon polarization versus $x$ at $Q^2=4$ GeV$^2$
from ref.\cite{BS95} (soft
gluon BS-s is dotted curve, hard gluon BS-h is dashed curve) and from
ref.\cite{BouBuc} (BS-i solid curve).
\bigbreak \no
{\bf Fig. 1.b} The gluon polarization versus $x$  at $Q^2=4$ GeV$^2$
from ref.\cite{GS96}
(gluon A GSA is solid curve, gluon B GSB is dashed curve, gluon C GSC is dotted
curve) and from ref.\cite{GRSV} (standard scenario GRSV, dotted-dashed curve).
\bigbreak \no
{\bf Fig. 2} The $Q^2$ dependence of $A_1^p(x,Q^2)$ as predicted from
ref.\cite{GRSV} at fixed values of $x$.
\bigbreak \no
{\bf Fig. 3.a} The double helicity asymmetry $A_{LL}^{\gamma}(\eta=0)$
at $\eta = 0$ versus $p_T^{\gamma}$
for ${\sqrt s} = 500$ GeV. The solid curves, which are the exact calculations using
BS-h (upper curves) and GRSV (lower curves), are compared with the dotted curves obtained using eq.(9).
\bigbreak \no
{\bf Fig. 3b} Same as Fig. 3a using BS-h (solid curve), BS-s (large dashed curve),
BS-i (dotted curve), GRSV (dashed-dotted curve) GSA, GSB and GSC (small dashed
curves).
\bigbreak \no
{\bf Fig. 4} Various predictions for the $\eta$ distribution of 
$A_{LL}^{\eta}$ at  $p_T^{\gamma} = 40$ GeV/c and for ${\sqrt s} = 500$ GeV
(curve labels as in Fig.3b). 
The statistical errors are calculated for $\Delta \eta =0.5$ per bin and for a 
$p_T^{\gamma}$ acceptance $\Delta p_T^{\gamma} = 10$ GeV/c.
\bigbreak \no
{\bf Fig. 5} The double helicity asymmetry $A_{LL}^{jet}(\eta=0)$
at $\eta = 0$ and ${\sqrt s} = 500$ GeV versus $p_T^{jet}$, for various
parametrizations (curve labels as in Fig.3b). 
The statistical errors are calculated for $\Delta \eta =1$ per bin and  a 
$p_T^{jet}$ acceptance $\Delta p_T^{jet} = 10$ GeV/c. 
\bigbreak \no
{\bf Fig. 6} Various predictions for the $\eta$ distribution of 
$A_{LL}^{jet}(\eta)$
at  $p_T^{jet} = 60$ GeV/c  and ${\sqrt s} = 500$ GeV
(curve labels as in Fig.3b). 
The statistical errors are calculated for $\Delta \eta =0.2$  and  a 
$p_T^{jet}$ acceptance $\Delta p_T^{jet} = 10$ GeV/c.
\bigbreak \no
{\bf Fig. 7a} Various predictions for the $y$ distribution of 
$A_{L}^{W^+}(y)$
at  ${\sqrt s} = 500$ GeV. BS-h (solid curve), GRSV (dashed-dotted curve),
GSA (dashed curve), GSC (dotted curve). 
The statistical errors are calculated for $\Delta \eta =0.2$.
\bigbreak \no
{\bf Fig. 7b} Same as Fig.7a for $W^-$ production
\bigbreak
\no
{\bf Fig. 8a}
 Various predictions for the $y$ distribution of 
$A_{LL}^{PV}(y)$ for ${W^+}$ production
at  ${\sqrt s} = 500$ GeV (curve labels as in Fig.7a).
\bigbreak \no
{\bf Fig. 8b} Same as Fig.8a for $W^-$ production.
\bigbreak \no
{\bf Fig. 9a}  Various predictions for the $y$ distribution of 
$A_{LL}^{PV}(y)$ for ${W^+}$ production in $pn$ collisions
at  ${\sqrt s} = 500$ GeV (curve labels and statistical errors as in Fig.7a).
\bigbreak \no
{\bf Fig. 9b} Same as Fig.9a for ${\bar A}_{LL}^{PV}(y)$.





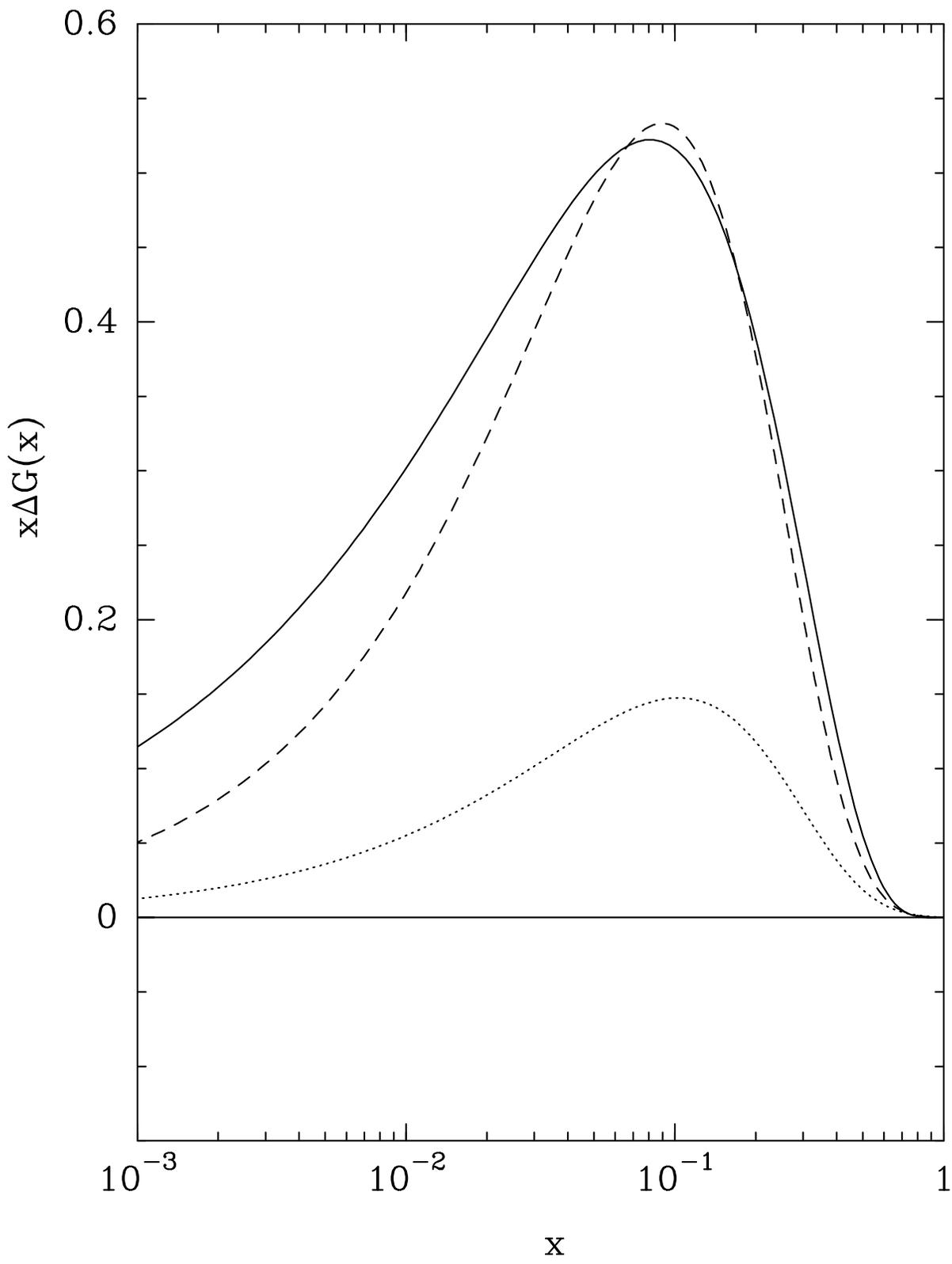
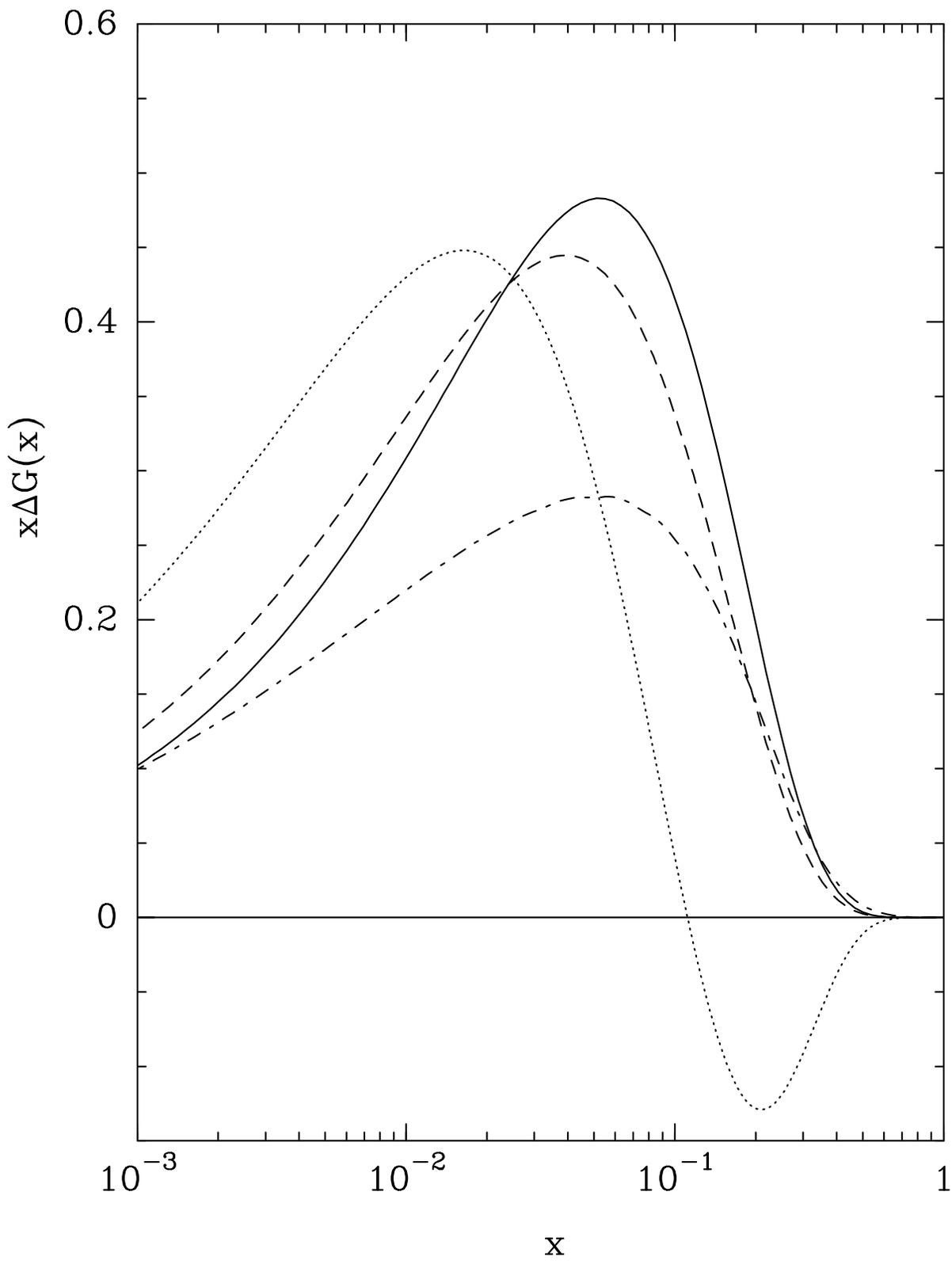
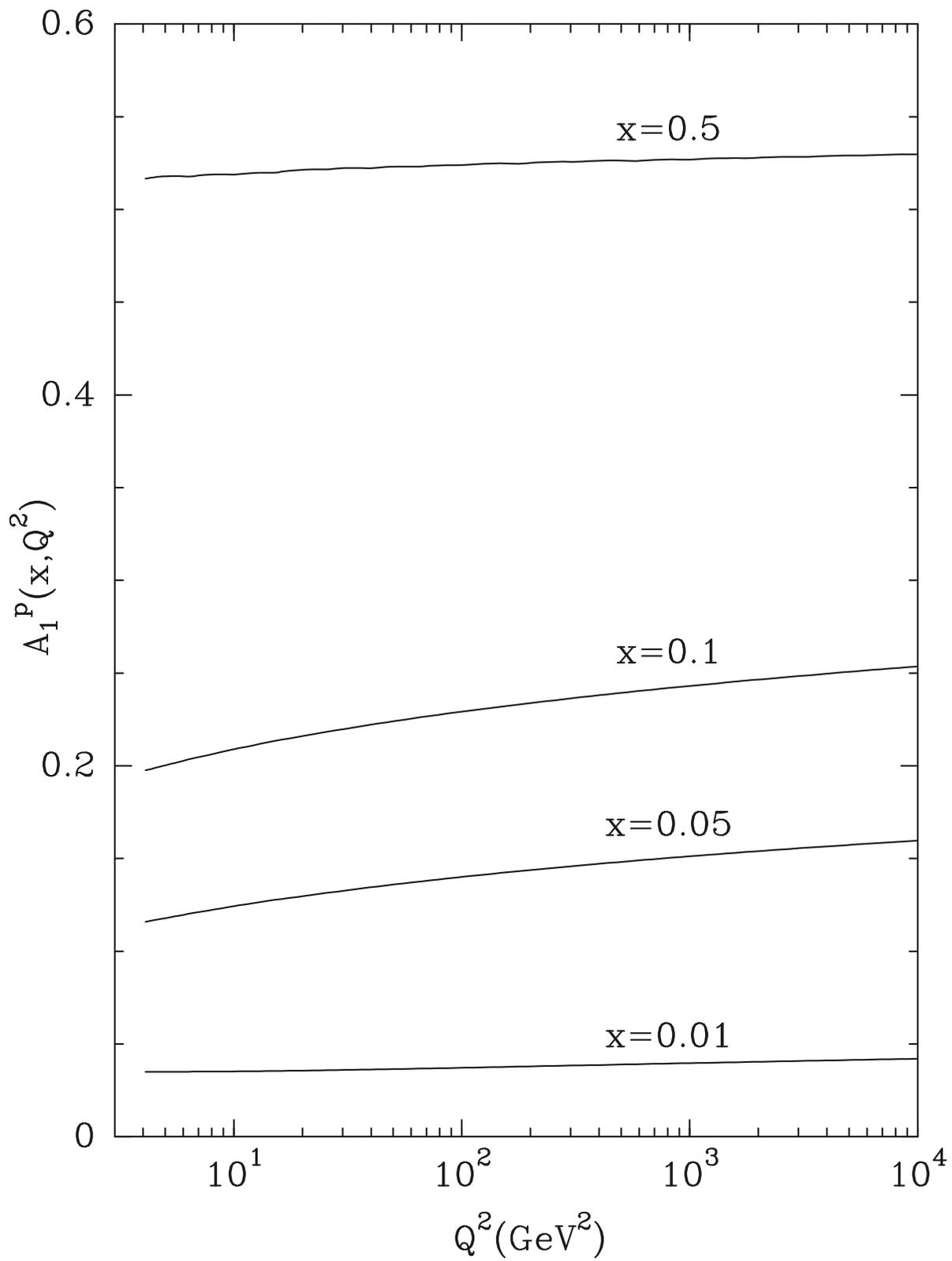
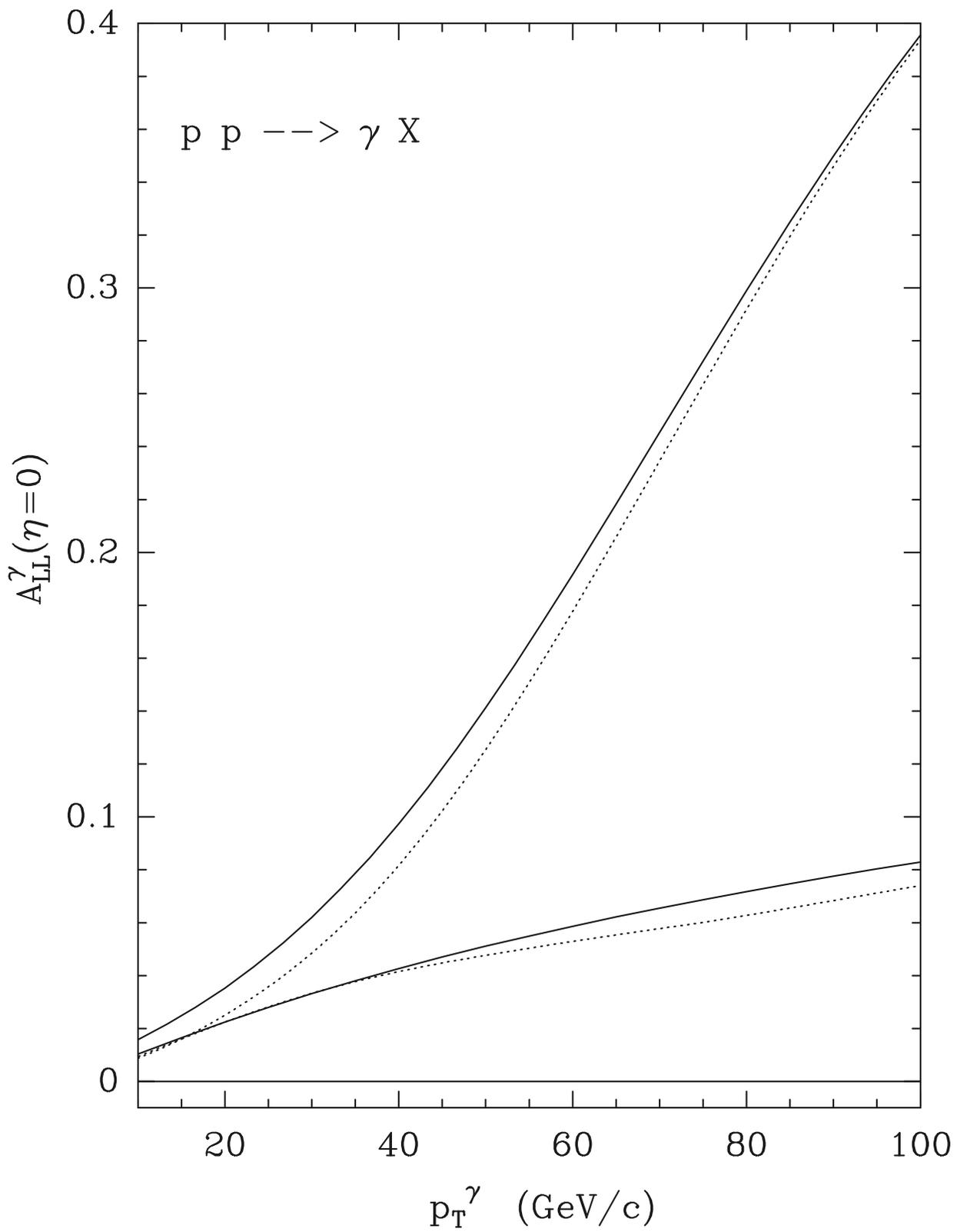
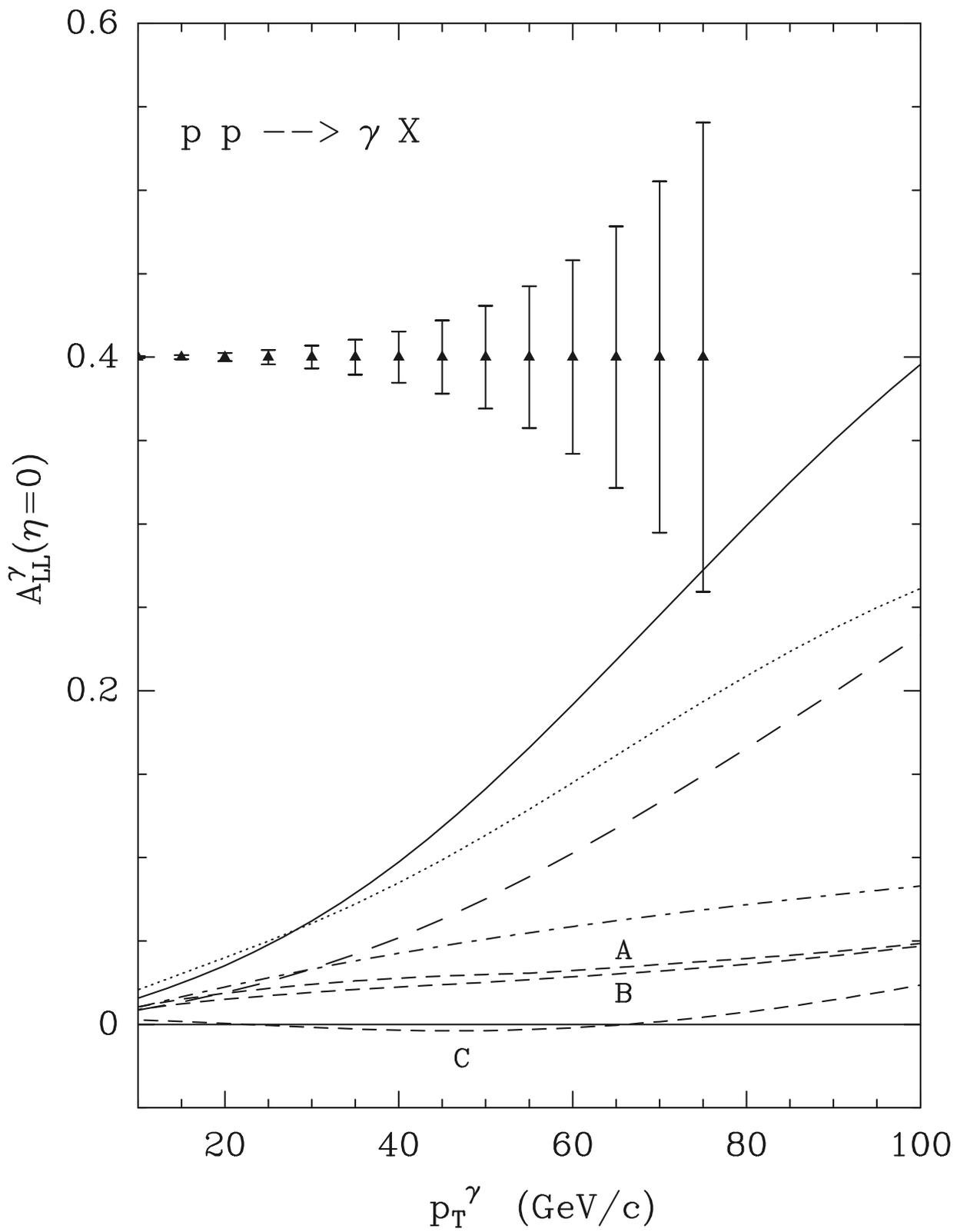
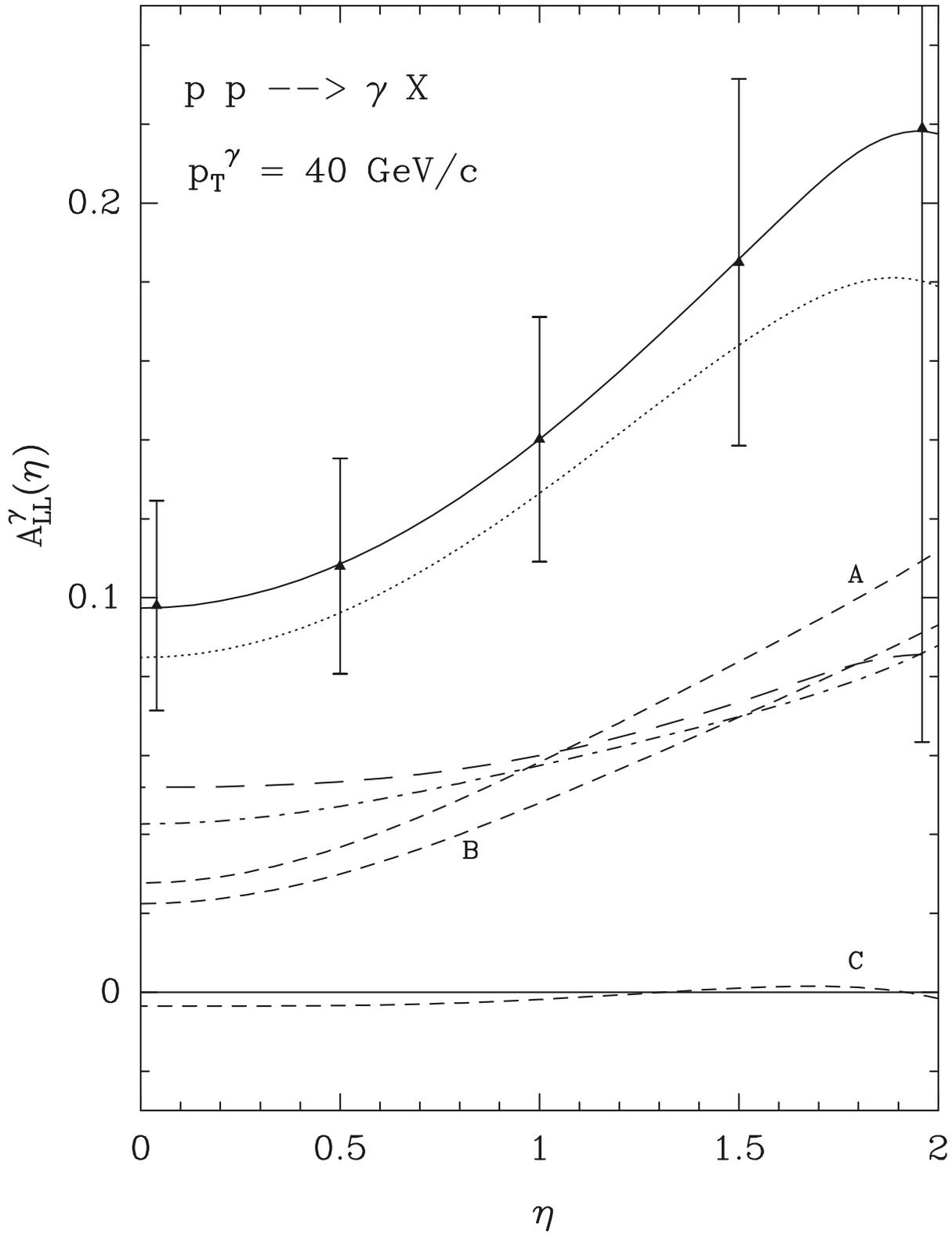
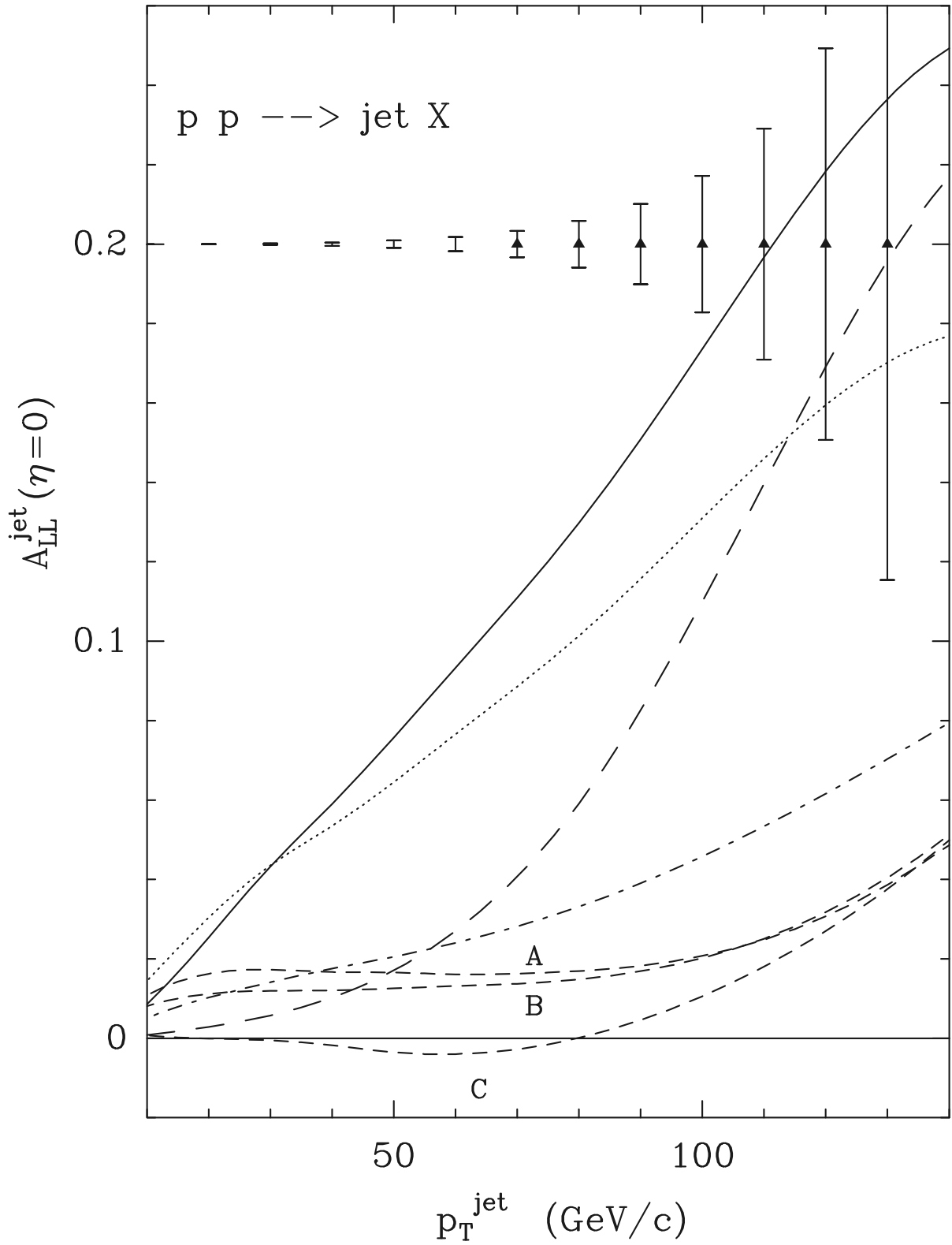
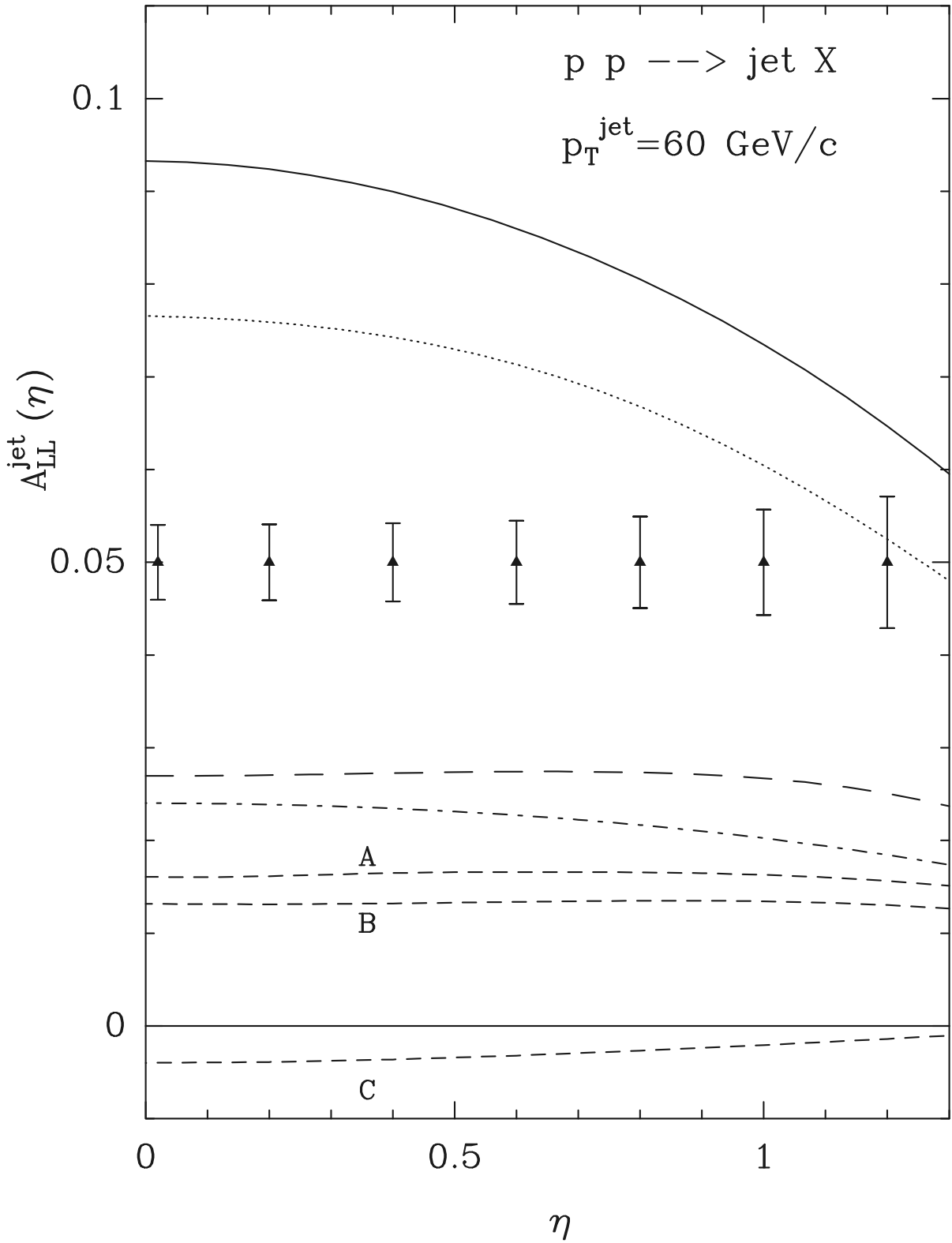
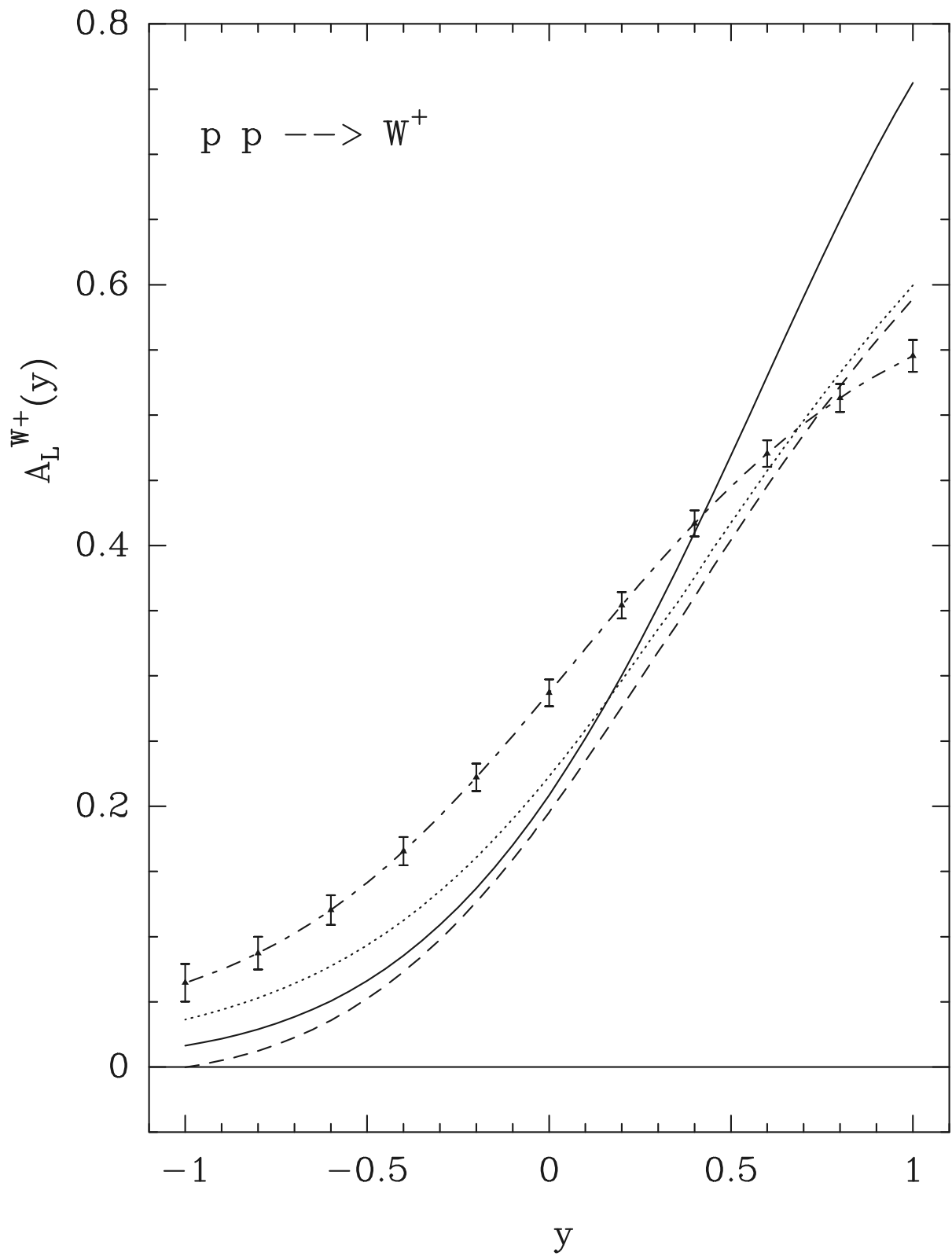
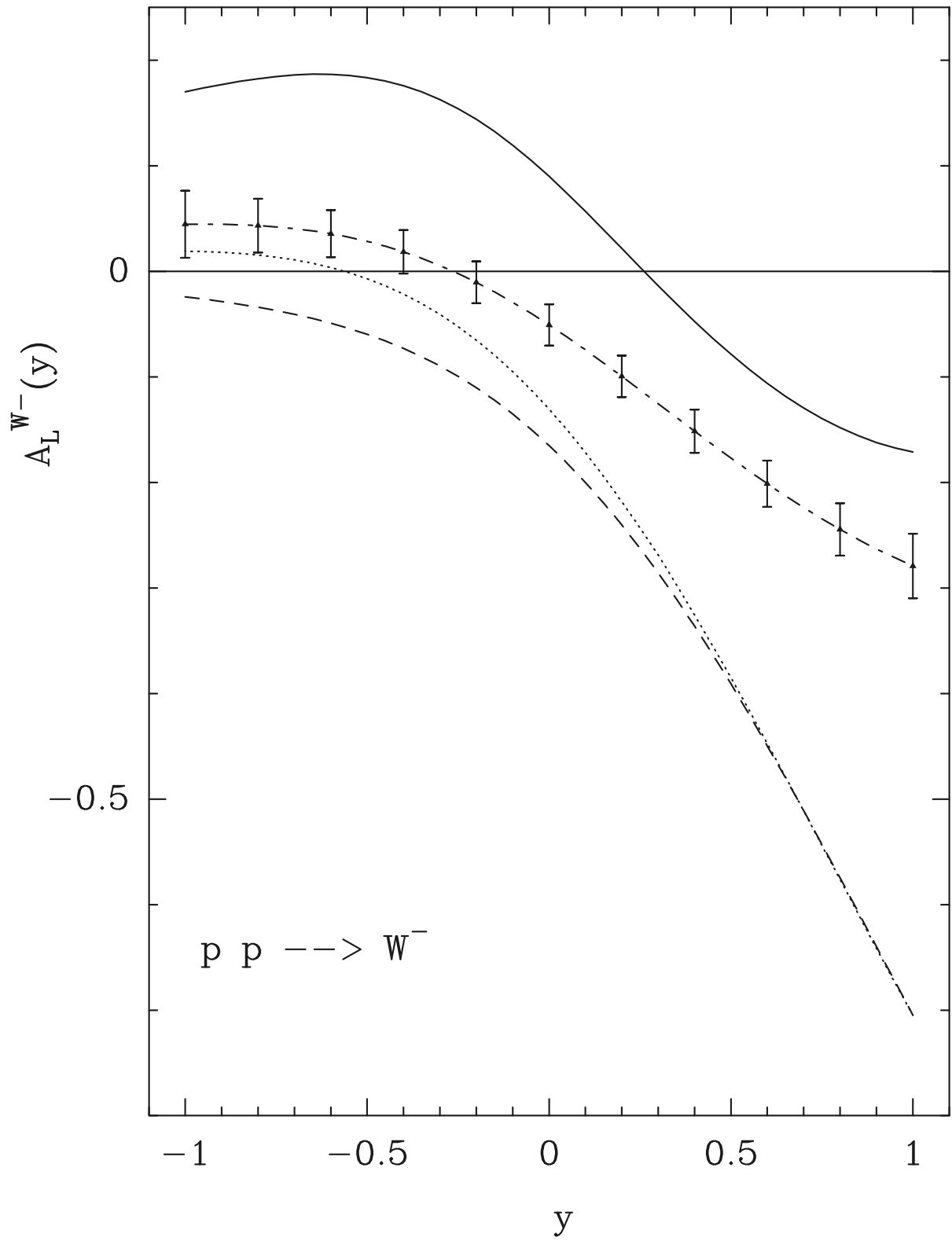
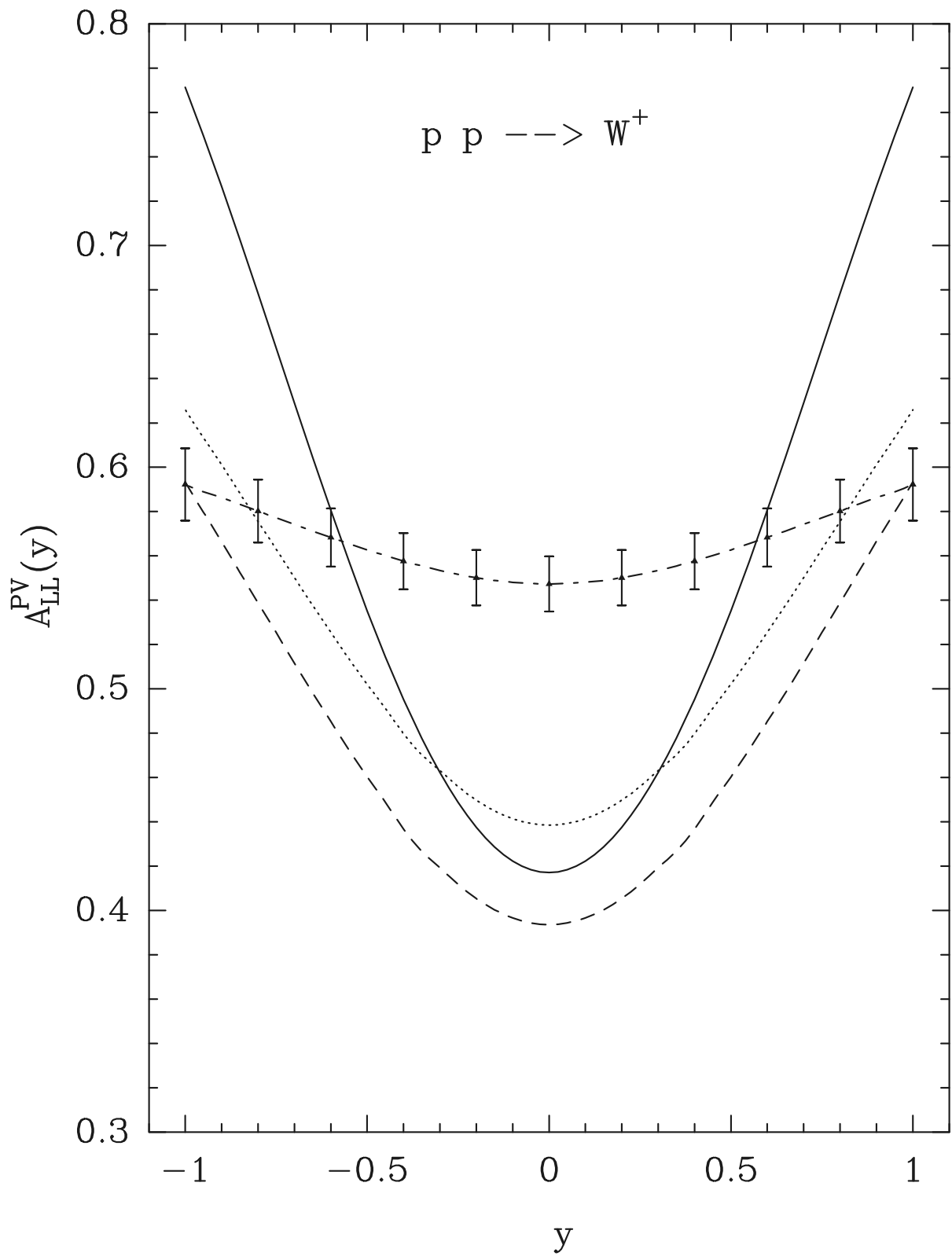
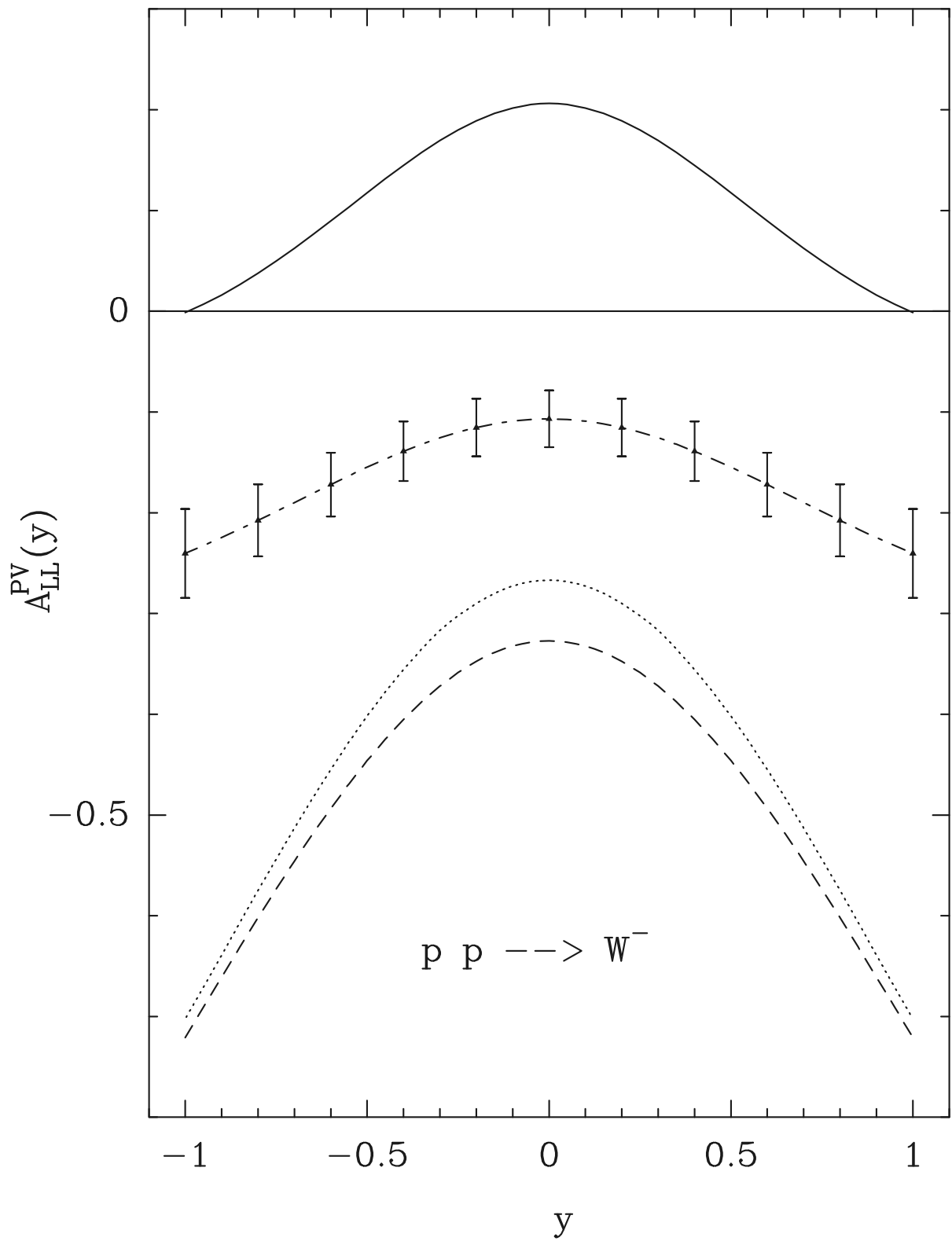
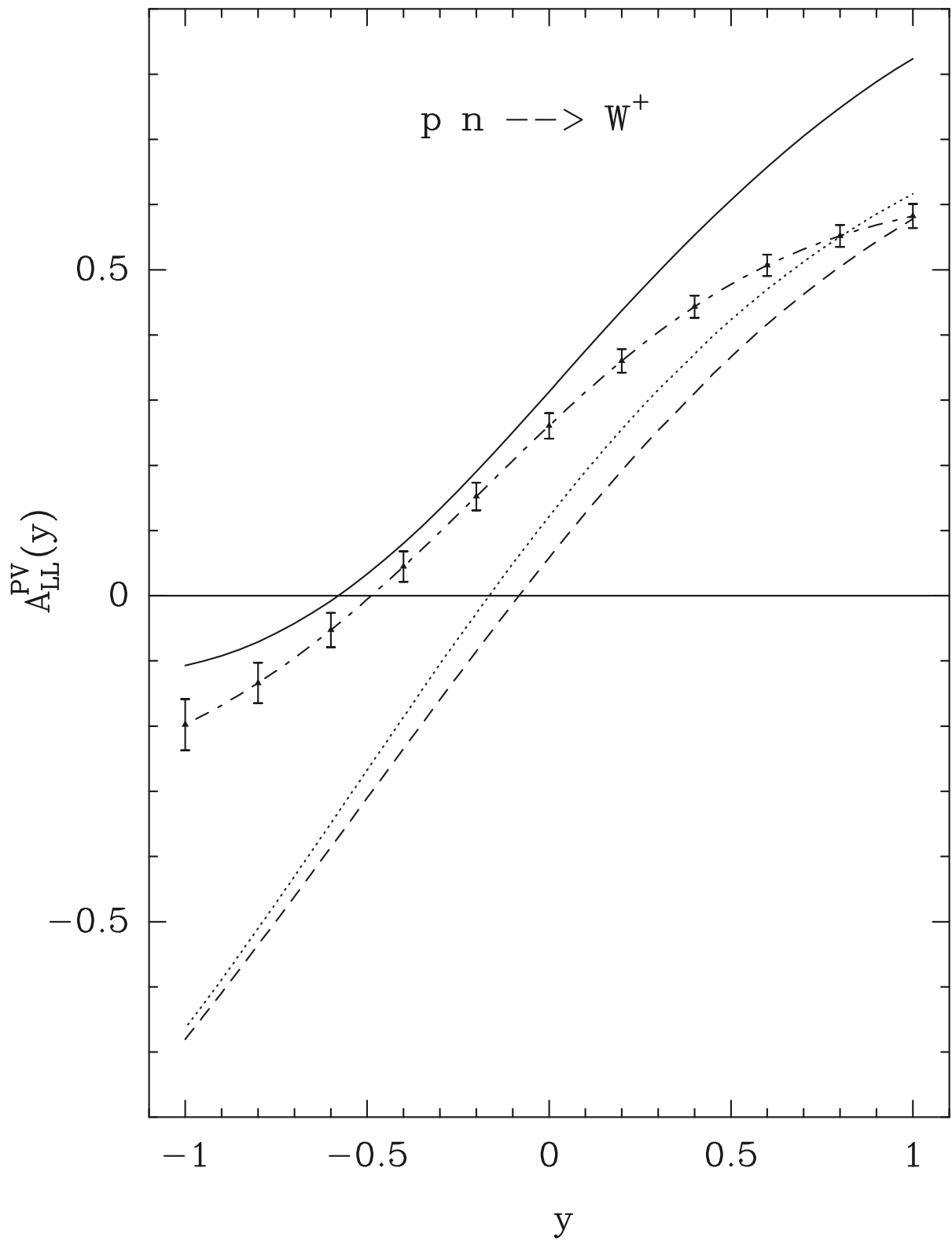
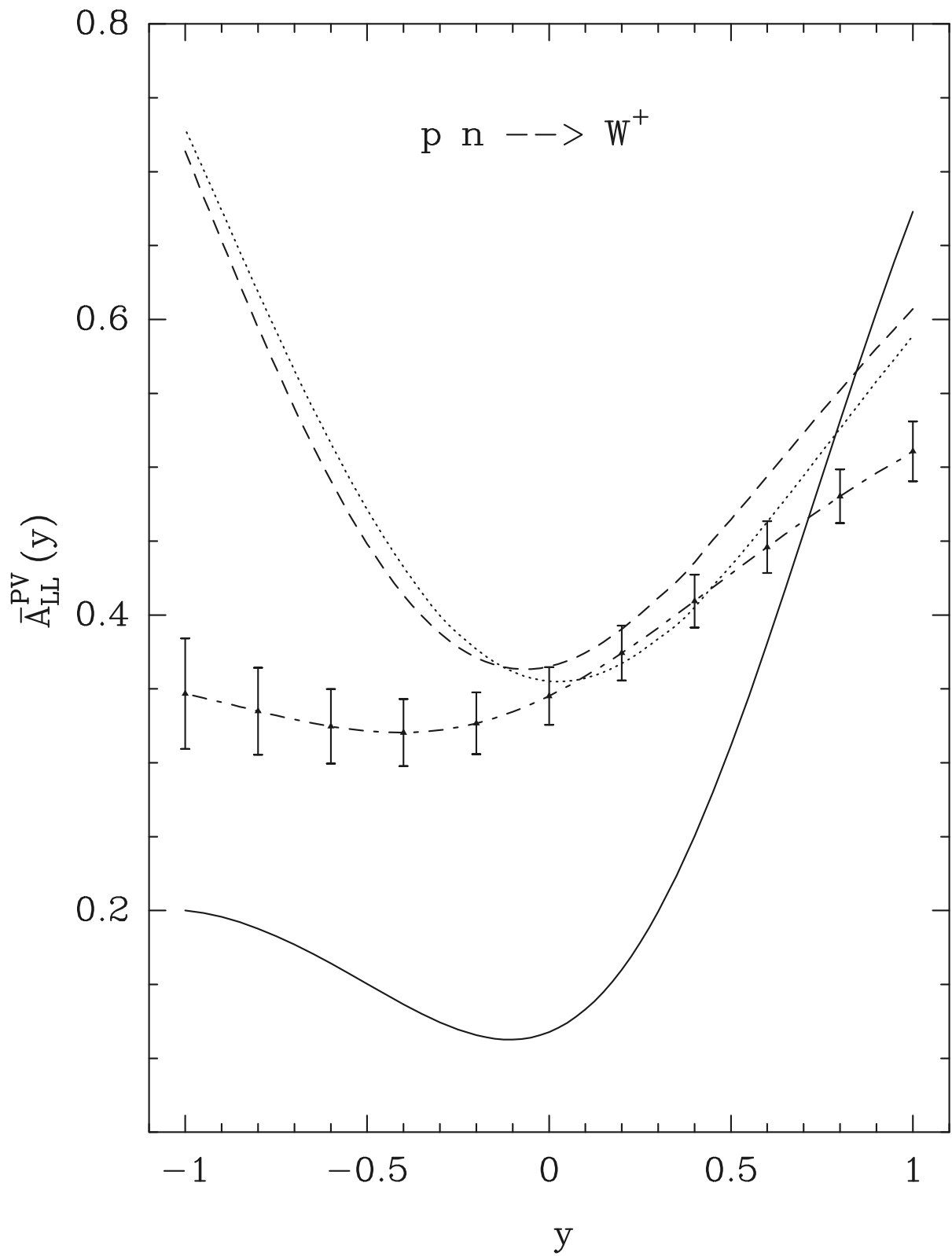
\end{document}